\documentclass{svproc}
\usepackage{graphicx}
\usepackage{mathtools}
\usepackage{siunitx}
\usepackage{amssymb}
\usepackage{pifont}
\usepackage{pbox}
\newcommand{\cmark}{\ding{51}}%
\usepackage{caption}
\usepackage{hyperref}
\let\OLDthebibliography\thebibliography
\renewcommand\thebibliography[1]{
  \OLDthebibliography{#1}
  \setlength{\parskip}{0pt}
  \setlength{\itemsep}{0pt plus 0.3ex}
}

\usepackage{url}

\begin{document}
\mainmatter              
\title{MM-Locate-News: Multimodal Focus Location Estimation in News}
\titlerunning{MM-Locate-News: Multimodal Focus Location Estimation in News}

\newcommand\orcid[1]{\href{https://orcid.org/#1}{\includegraphics[height=\fontcharht\font`A]{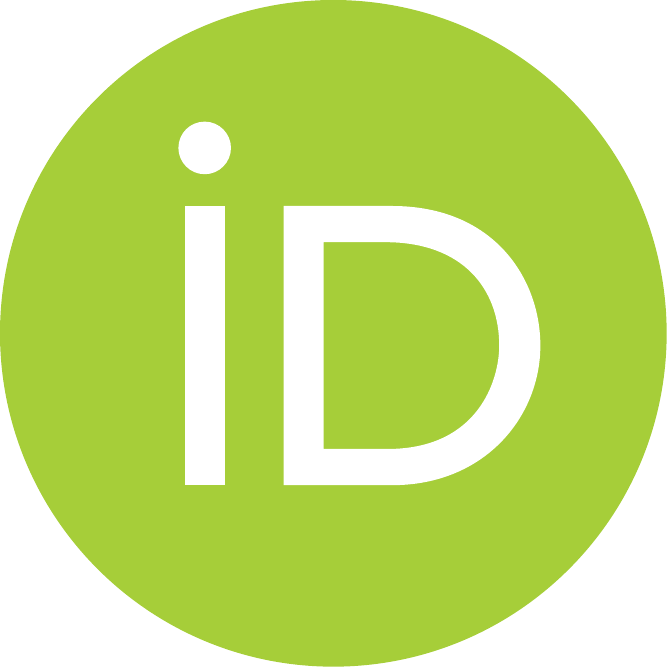}}}

\author{Golsa Tahmasebzadeh\inst{1,2}\orcid{0000-0003-1084-5552} \and  
Eric M\"uller-Budack\inst{1,2}\orcid{0000-0002-6802-1241} \and 
Sherzod Hakimov\inst{3}\orcid{0000-0002-7421-6213} \and
Ralph Ewerth\inst{1,2}\orcid{0000-0003-0918-6297}}
\authorrunning{G. Tahmasebzadeh et al.}
\institute{TIB--Leibniz Information Centre for Science and Technology, Hannover, Germany \and
L3S Research Center, Leibniz University Hannover, Germany
\and
University of Postdam, Germany \\
\email{\{golsa.tahmasebzadeh, eric.mueller, ralph.ewerth\}@tib.eu, sherzod.hakimov@uni-potsdam.de}}

%

\maketitle              

\begin{abstract}
The consumption of news has changed significantly as the Web has become the most influential medium for information. To analyze and contextualize the large amount of news published every day, the geographic focus of an article is an important aspect in order to enable content-based news retrieval. 
There are methods and datasets for geolocation estimation from text or photos, but they are typically considered as separate tasks. However, the photo might lack geographical cues and text can include multiple locations, making it challenging to recognize the focus location using a single modality. In this paper, a novel dataset called Multimodal Focus Location of News (MM-Locate-News) is introduced. We evaluate state-of-the-art methods on the new benchmark dataset and suggest novel models to predict the focus location of news using both textual and image content. The experimental results show that the multimodal model outperforms unimodal models. 

\end{abstract}
\section{Introduction}\label{sec:intro}
\begin{figure}[t]
  \includegraphics[trim=0cm 0.2cm 0cm 0cm,clip,width=\textwidth]{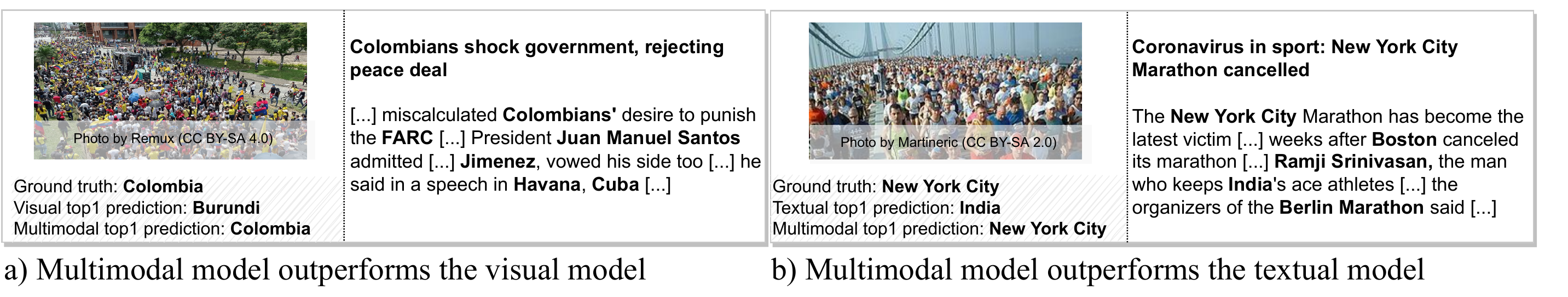}  \caption[width=0.5\textwidth]{  Samples from the MM-Locate-News dataset. (a) The proposed visual model fails due to lack of visual geo-representative clues. (b) The proposed textual model fails due to various locations in the text. In both cases, the combination of visual and textual features helps specify the correct location. Images are replaced with similar ones due to license restrictions.}
  \label{fig:sample_outputs}
\end{figure}

With the rapidly growing amount of information on the Web and social media, news articles are increasingly conveyed in multimodal form; apart from text, images and videos have become inseparable from news to report on events. 
Besides, there is an increasing research interest in the geographical content of news in fields like information retrieval or Web science. 
Although a news article might mention many locations it is important to determine the location of the main story. The concept of \textit{focus} location is a piece of metadata which indicates the geographic focus of a news article~\cite{clavin}. 
The focus locations are vital, for instance, to identify the location of a natural disaster for crisis management or to analyse political or social events across the world.
Even if the news article focuses on one particular event, the text usually mentions various locations which are not the focus locations~(see Figure~\ref{fig:sample_outputs}b). 
On the other hand, a news image alone usually does not include sufficient geo-representative features to identify the focus location reliably~(see Figure~\ref{fig:sample_outputs}a).
Therefore, approaches for multimodal geolocation estimation are required to bridge these gaps. 

Most of previous approaches for geolocation estimation depend on either text \cite{clavin,mordecai,profile,kulkarni2021multi} or image~\cite{mvmf,kim2017learned,hierarchical}.
Text-based approaches assign geo-coordinates to the focus location at the document level~\cite{clavin,andogah2012every} or based on specific events~\cite{mordecai,profile}. 
On the other hand, the majority of image-based methods focus on locating photos from certain environments such as cities~\cite{kim2017learned,DBLP:conf/eccv/GordoARL16}. In recent years, several deep learning approaches have been introduced for geolocation estimation at global scale without any prior assumptions~\cite{DBLP:conf/cvpr/LinBH13,hierarchical,cplanet,googlelandmarkdataset}.
However, information from image and text are required for a more robust focus location estimation in multimodal news, as the examples in Figure~\ref{fig:sample_outputs} illustrate. But, so far, only a few methods have considered multiple modalities for geolocation estimation~\cite{kordopatis2017geotagging,placing_language_model}.
Overall, these methods suffer from two major limitations:
(1)~They do not make use of recent multimodal transformer models~\cite{clip} or other unimodal state-of-the-art approaches to extract rich textual~(e.g., \cite{bert}) and visual information~(e.g., \cite{hierarchical});
(2)~Although there are various large-scale datasets for text-based geolocation estimation of news at document level~(e.g., Local-Global Lexicon~(LGL)~\cite{lgl},  GeoVirus~\cite{geovirus}), or based on events~(e.g., Atrocity Event Data~\cite{event_data}, political news from New York Times~\cite{profile}), they do not provide the accompanying images and thus do not allow for multimodal focus location. 

Some multimodal datasets, such as the \textit{MediaEval Placing Task} benchmark datasets~\cite{mediaeval2}, WikiSatNet~\cite{BurakEvan}, or \textit{Multiple Languages and Modalities (MLM)}~\cite{mlm}, contain image-text pairs with the corresponding geo-coordinates, but do not cover news articles and consequently do not address the related challenges such as multiple entities in the text or the lack of geo-representative features in the images.
To the best of our knowledge, \textit{BreakingNews}~\cite{breakingnews} is the biggest dataset of multimodal news articles with corresponding location information. However, the locations are primarily extracted based on the \textit{Rich Site Summary}~(RSS) and, if not available, on heuristics such as the publisher location or story text~\cite{breakingnews}. As a result, each document is provided with a number of potential locations that do not necessarily correspond to the focus location, are inaccurate, or even wrong in some cases. 
In summary, there is a clear need for multimodal datasets of news articles with image and text and high-quality focus location labels. In addition, a multimodal approach is required which benefits from state-of-the-art approaches to provide rich visual and textual features for news documents.

In this paper, we address the task of focus location estimation; specifically, we target the gap between lack of visual geo-representative clues in news images and multiple location mentions in the body text of news. 
%
Our contributions are summarized as follows. 
(1)~We present the \textit{MM-Locate-News (Multimodal Focus Location of News)} dataset that consists of \num{6395} news articles covering \num{237} cities and \num{152} countries across all continents as well as multiple domains such as \textit{health}, 
\textit{environment}, and \textit{politics}. 
The dataset is collected in a weakly-supervised manner, and multiple data cleaning steps are applied to remove articles with potential inaccurate geolocation information. The acquired dataset addresses drawbacks of other datasets such as BreakingNews~\cite{breakingnews} as it considers multimodal content of news to label the corresponding location. 
Furthermore, we provide a test set of \num{591} manually annotated news articles. 
\noindent(2)~We propose several neural network models that, unlike previous work~\cite{kordopatis2017geotagging,placing_language_model,DBLP:conf/www/CrandallBHK09,breakingnews,DBLP:conf/sigir/SerdyukovMZ09}, exploit state-of-the-art image-based~\cite{hierarchical,zhouplaces}, text-based~\cite{bert}, and multimodal approaches~\cite{clip} to extract embeddings 
for multimodal geolocation estimation of news. 
\noindent(3)~Experimental results and comparisons to existing solutions on the proposed dataset and on BreakingNews~\cite{breakingnews} demonstrate the feasibility of the proposed approach.

The remainder of the paper is structured as follows. Section~\ref{sec:mmg_news} describes the dataset acquisition, while the proposed model for multimodal geolocation estimation is presented in Section~\ref{sec:baseline_models}. We discuss the experimental results in Section~\ref{sec:results}. Section~\ref{sec:conc} concludes the paper and outlines potential directions for future work.

\section{MM-Locate-News Dataset}
\label{sec:mmg_news}
\begin{figure}[t]
\centering
  \includegraphics[width=0.8\linewidth]{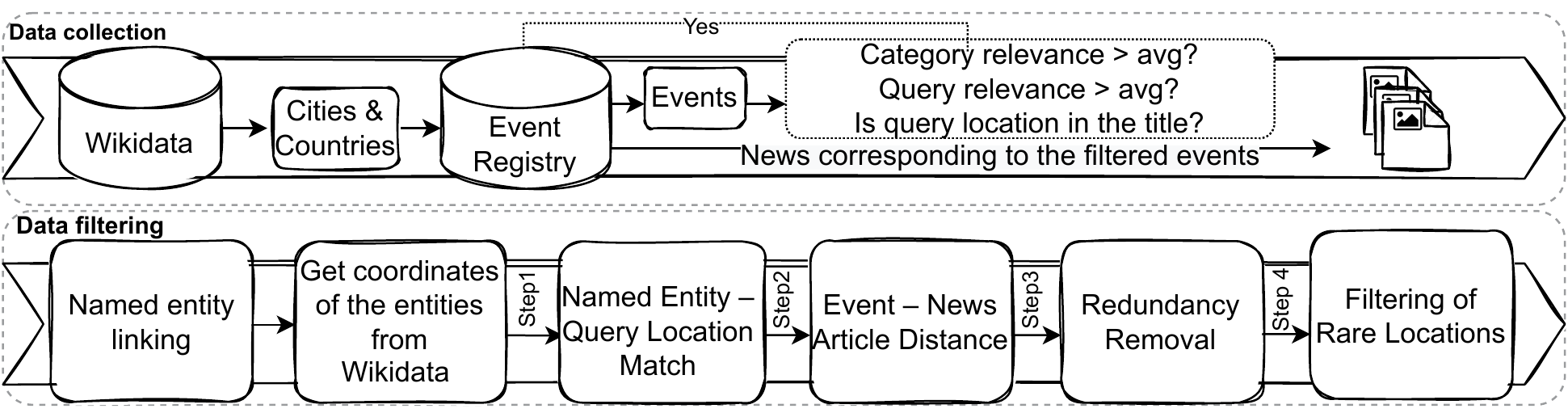}
  \caption{MM-Locate-News data collection and filtering steps.}
  \label{fig:data_collection}
\end{figure}

In this section, we present our novel dataset called \textit{Multimodal Focus Location of News (MM-Locate-News)}\footnote{Source code: \url{https://github.com/TIBHannover/mm-locate-news}}. 
As discussed in Section~\ref{sec:intro}, existing datasets for geolocation estimation are either not related to news~\cite{mediaeval2,BurakEvan,DBLP:journals/cacm/VrandecicK14}, do not provide images along with news texts~\cite{lgl,geovirus}, or contain unreliable labels for locations extracted from the news feed~\cite{breakingnews}.
In contrast, MM-Locate-News includes image-text pairs of news and the location focus at different geospatial resolutions, such as city, country, and continent. 
In the sequel, data collection and cleaning steps~(Figure~\ref{fig:data_collection}) as well as annotation process and dataset statistics are presented.

\textbf{Data Collection}: The dataset has been collected in a weakly-supervised fashion. To cover a variety of locations from all six continents, we extract all countries, capitals, highly populated cities~(minimum population of \num{500000}; minimum area of $200\,\textnormal{km}^2$), and medium populated cities~(population between \num{20000} and \num{500000}; area between $100\,\textnormal{km}^2$ and $200\,\textnormal{km}^2$) from Wikidata~\cite{DBLP:journals/cacm/VrandecicK14}. 
For each location, we query \textit{EventRegistry}\footnote{\url{http://eventregistry.org/}} for events between 2016 and 2020 from the following categories: \textit{sports}, \textit{business}, \textit{environment}, \textit{society}, \textit{health}, and \textit{politics}. 
Note that \textit{EventRegistry} automatically clusters news articles reporting on the same (or similar) events and that the news title of the cluster centroid represents the event name. To ensure the quality, we filter out events that do not include the location in their name or when their \textit{category relevance} and \textit{query relevance} scores, provided by \textit{EventRegistry}, are below the average scores of all events per query location. The intuition behind this step is that an event with a location mentioned in its name more likely provides news articles focusing on the queried location. 
Finally, we extract all news articles from the remaining event clusters.

\textbf{Data Filtering}: We apply the following steps to remove irrelevant samples.\\
\noindent\textit{1) Named Entity -- Query Location Match:} We assume that an article is related to a query location if it is geographically close to at least one named entity. 
%
Following related work~\cite{DBLP:journals/ijmir/Muller-BudackTD21}, we extract the named entities using \emph{spaCy}~\cite{spacy} and use \textit{Wikifier}~\cite{wikifier} to link them to \textit{Wikidata} for disambiguation. 
We extract \textit{coordinate location}~(\textit{Wikidata Property P625}), which is available primarily for locations~(e.g., landmarks, cities, or countries). 
For persons, we extract the \textit{place of birth}~(\textit{Wikidata Property P19}) as they likely act in the respective country~(or even city). %
We compute the Great Circle Distance~(GCD) between the geographical coordinates of the query location and the extracted entity locations.
We keep news articles that include at least one named entity whose GCD from the query location is smaller than $\sqrt[k]{a}$ where $a$ is the area~(\textit{Wikidata Property P2046}) of the query location, and $k$ is a hyperparameter as defined in Section~\ref{sec:results}.\\
\noindent\textit{2) Event -- News Article Distance:} Each news article in \textit{EventRegistry} is assigned a similarity measure that represents the closeness to an event. 
We discard articles with a lower similarity than the average similarity of all articles of the same cluster to keep the news articles that are most related to the respective event.\\
\noindent\textit{3) Redundancy Removal:} We compute the similarity between news articles using TF-IDF vectors (Term Frequency; Inverse Document Frequency) and discard one of the articles when the similarity is higher than \num{0.5} to remove redundancy.\\
\noindent\textit{4) Filtering of Rare Locations:} After applying the filtering step 1-3, we remove rare locations~(and corresponding articles) with less than five articles as they contain too few articles for training. 

\textbf{Dataset Statistics}:
In total, we queried \num{853} locations and extracted \num{13143} news articles. 
After the data cleaning steps, we end up with \num{6395} news articles for \num{389} locations (237 cities and 152 countries).
We divided the \emph{MM-Locate-News} dataset into train, validation and test data splits by equally distributing news articles among locations as given in Table~\ref{tab:data_world}, yielding approximately 80:10:10 splits (see Figure~\ref{fig:sample_outputs} for samples from the dataset).

\begin{table}[t]
\begin{center}
\parbox{.48\linewidth}{
\caption{Distribution of train, validation and test samples in MM-Locate-News among continents~(AF: Africa, SA: South America, EU: Europe, AS: Asia, NA: North America, OC: Oceania)}
\label{tab:data_world}
\resizebox{0.48\columnwidth}{!}{
\begin{tabular}{  l | c  c  c  c  c  c | c } \hline  
& \textbf{AF} & \textbf{SA} & \textbf{EU} & \textbf{AS} & \textbf{NA} & \textbf{OC} & \textbf{Total} \\ \hline  
\textbf{Train}& 854& 216& \num{1604}& \num{1842}& 589& 161& \num{5266} \\  
\textbf{Val}& 93& 24& 147& 160& 88& 23& 535 \\  
\textbf{Test}& 84& 29& 179& 202& 71& 26& 591 \\  \hline  
\end{tabular} 
}
}
\hfill
\parbox{.48\linewidth}{
\caption{Manual annotation criteria~(C) for the MM-Locate-News test set variants~(T). Answers “yes”, “no" or “unsure" are denoted as “-”, while “u" and “\cmark" denote “unsure" and “yes".}
\label{tab:test_data} 
\resizebox{0.48\columnwidth}{!}{
\begin{tabular}{  l |  c  c  c  c } \hline  
&  \textbf{T1} & \textbf{T2} & \textbf{T3} \\ \hline 
\textbf{C1:} Image depicts query location & - & \cmark& u \\  \hline
\textbf{C2:} Text focuses on query location & \cmark& \cmark& \cmark \\ 
\hline
\textbf{C3:} Image and text conceptually related& -& -& \cmark \\ \hline 
\textbf{Number of samples}& 591& 65& 154 \\  \hline  
\end{tabular} 
}
}
\end{center} 
\end{table}

\textbf{Data Annotation}: The test split of the dataset is manually annotated. Users annotated a given news article along with its image and the query location to provide “yes", “no", or “unsure" labels to three criteria~(\textit{C1 -- C3}) given in Table~\ref{tab:test_data}. Different criteria depending on the answers are turned into a different variant of the test data to evaluate the geolocation estimation models.
In the \textit{T1} version, the text focuses on the query location, and in the \textit{T2} both image and text represent the query location. Since it was difficult to find images where the query location is shown, we made the \textit{T3} version where the annotators were not certain about whether the image shows the location. Thus, in cases where the text focuses on the location and the image and text are related, we assume that the image also shows the location.

\textbf{Annotator Agreement}: A total of three users annotated the test set, where two users annotated each sample. 
The inter-coder agreement for the criteria \textit{C1}, \textit{C2}, and \textit{C3} is \num{0.44}, \num{0.38}, and \num{0.55}, respectively (according to Krippendorff's alpha~\cite{krippendorff2011computing}).
Despite relatively moderate agreement scores, we noticed that the agreement in percent is  quite high: \textit{C2} and \textit{C3} are \num{80}\% and \textit{C1} is \num{66.6}\%. This is caused by the annnotators' bias towards the answer “yes" for all criteria.

\section{Multimodal Focus Location Estimation}\label{sec:baseline_models}

We define focus location estimation as a classification problem where for each article, i.e., image-text pair, the total number of $n$~query locations (country or city) are considered as target classes. 
The $n$-dimensional one-hot encoded ground truth vector~\mbox{$\mathbf{y} = \langle y_1, y_2, \dots , y_n \rangle \in \{0, 1\}^n$} represents the query location. 
We extract textual and visual features as follows.

\textbf{Visual Features}: The visual \textit{Scene} descriptor (representing a place in a general sense) is based on ResNet-152 model~\cite{resnet} to recognize 365 places (pre-trained on the Places365 dataset~\cite{zhouplaces}). 
The \textit{Location} descriptor, $base(M, f^*)$, is taken from a state-of-the-art photo geolocation estimation approach~\cite{hierarchical}. 
The \textit{Object} descriptor utilizes the ResNet-152 model~\cite{resnet} pre-trained on the Image\-Net dataset~\cite{deng2009imagenet}. 
The \textit{CLIP$_i$} descriptor is extracted using CLIP (Contrastive Language-Image Pretraining)~\cite{clip} image encoder.

\textbf{Textual Features}: A pre-trained BERT~(Bidirectional Encoder Representations from Transformers)~\cite{bert} model is used to extract two distinct feature vectors. 
The first one represents the whole body text (BERT-Body), while the other refers to mentions of named entities and averages the individual word embeddings of all named entities (BERT-Entities).

\begin{figure}[t]
\centering
  \includegraphics[width=0.9\linewidth]{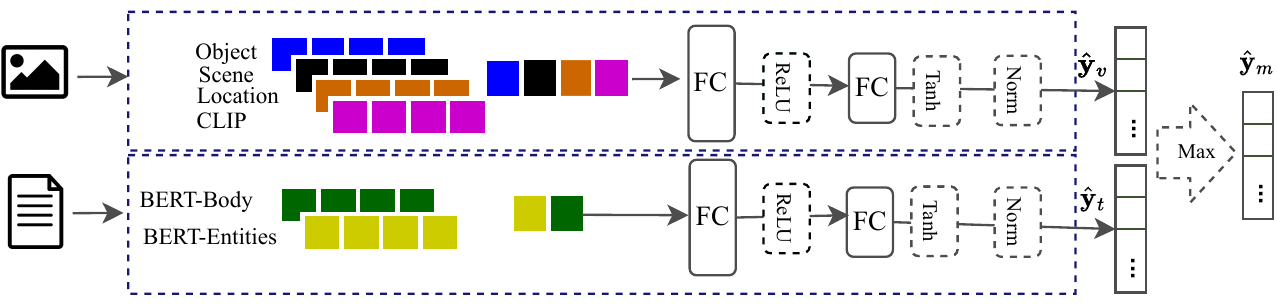}
  \caption{The model architecture for multimodal focus location estimation in news.}
  \label{fig:models}
\end{figure}

\textbf{Multimodal Architecture}: As shown in Figure~\ref{fig:models}, the textual and visual embeddings are concatenated and passed to the fully connected (FC) layers with an output size of 1024, followed by a \textit{ReLU} layer. 
Next, the outputs are fed to another \textit{FC} layer followed by a $\tanh(u_i^l)$ layer. 
The norm function with clamp $\min=10^{-12}$ is applied to extract the visual~$\hat{\mathbf{y}}_v$ and textual vector~$\hat{\mathbf{y}}_t$ with size of $n=389$ (number of locations). 
We obtain the multimodal output~$\hat{\mathbf{y}}_m$ using the maximum probabilities of the visual~$\hat{\mathbf{y}}_v$ and textual~$\hat{\mathbf{y}}_t$ outputs.

\textbf{Network Optimization}:
During training, the models are optimized as follows. 
For unimodal models, the cross-entropy loss between the ground truth one-hot encoded vector~$\mathbf{y}$ to one~($\hat{\mathbf{y}}_v$ or $\hat{\mathbf{y}}_t$) is optimized. For multimodal models, the same procedure is applied, but on both vectors $\hat{\mathbf{y}}_v$ and $\hat{\mathbf{y}}_t$. 
We freeze the weights of the visual models during training.

\section{Experimental Setup and Results}\label{sec:results}
In this section, the evaluation setup~(Section~\ref{sec:eval_setup}) and experimental results including a comparison with state-of-the-art approaches on the \emph{MM-Locate-News}~(Section~\ref{sec:res_mufl}) and \textit{BreakingNews}~\cite{breakingnews}~(Section~\ref{sec:res_breakingnews}) datasets are reported.

\subsection{Evaluation Setup}
\label{sec:eval_setup}

\textbf{Evaluation Metrics}:
We use the geographical location of the class with highest probability in the network output~($\hat{\mathbf{y}}_v$, $\hat{\mathbf{y}}_t$, or $\hat{\mathbf{y}}_m$) as the prediction and measure the Great Circle Distance (GCD) to the ground truth location. 
As suggested by Hays and Efros~\cite{Hays2008}, the models are evaluated with respect to the percentage of samples predicted within different accuracy levels: \textit{city}, \textit{region}, \textit{country}, and \textit{continent} with maximum tolerable GCD to the ground location of \num{25}, \num{200}, \num{750}, and \num{2500} kilometers, respectively. 
For the evaluation on BreakingNews~(Section~\ref{sec:res_breakingnews}), we use mean and median GCD metrics as suggested by Ramisa et al.~\cite{breakingnews}.

\textbf{Hyperparameters}: 
We use the Adam optimizer, learning rate of \num{e-4}, and batch size of \num{128} for optimization. We train the models for 500 epochs and perform validation after every epoch. The best model is used based on the highest mean accuracy in city and region GCD thresholds on the validation set. We set the parameter for data filtering to $k=\num{6}$ (Section~\ref{sec:mmg_news}) based on an empirical evaluation of a small subset (150 samples) of the dataset.

\textbf{Compared Systems}: We evaluate different combinations of the proposed model based on the feature modalities. We also compare against two popular text-based methods \textit{Cliff-clavin}~\cite{clavin}, \textit{Mordecai}~\cite{mordecai} and one image-based state-of-the-art model 
(\textit{ISNs: Individual Scene Networks}~\cite{hierarchical}). 

\subsection{Experimental Results on MM-Locate-News}
\label{sec:res_mufl}
The results are reported in Table~\ref{table:eval} and 
discussed below.

\begin{table}[t]
\begin{center}
\caption{Accuracy~[\%] of focus location estimates for Cliff-clavin~\cite{clavin}, Mordecai~\cite{mordecai}, ISNs~\cite{hierarchical} and baselines on different test variants of MM-Locate-News (best results for each modality is highlighted per GCD accuracy levels). Text features: BERT-Body (B-Bd), BERT-Entities (B-Et). Visual Features: Location (Lo), Object (Ob), Scene (Sc), CLIP$_i$. GCD accuracy levels: City (CI, max. 25\,km GCD to the ground truth location), Region (RE, max. 200\,km), Country (CR, max. 750\,km), Continent (CT, max. 2500\,km)}
\label{table:eval}
\resizebox{\columnwidth}{!}{%
\begin{tabular}{  l | c | c  c  c  c | c  c  c  c | c  c  c  c } \hline

&  &\multicolumn{4}{c|}{T1} & \multicolumn{4}{c|}{T2} & \multicolumn{4}{c}{T3}  \\
  \cline{3-14}
\textbf{Approach} &\textbf{Modality} & \textbf{CI} & \textbf{RE} & \textbf{CR} & \textbf{CT} & \textbf{CI} & \textbf{RE} & \textbf{CR} & \textbf{CT} & \textbf{CI} & \textbf{RE} & \textbf{CR} & \textbf{CT} \\ \hline 
Mordecai~\cite{mordecai} (only country-level)&Textual & - & - & \textbf{74.1} & 82.9 & - & - & 72.3 & 84.6 & - & - & \textbf{72.7} & 81.8 \\ 
Cliff-clavin~\cite{clavin} 
& Textual & 36.9 & 53.1 & 71.2 & \textbf{86.5} & 38.5 & 56.9 & 66.2 & 87.7 & 33.1 & \textbf{48.7} & \textbf{72.7} & \textbf{85.1} \\  
B-Bd &Textual & 22.8 & 27.4  & 41.1  & 68.4 &33.8 & 35.4  & 49.2  & 70.8 &19.5 & 23.4  & 40.9  & 63.6   \\
B-Et  &Textual &  \textbf{48.1} & \textbf{53.5}  & 66.3  & 79.5 &58.5 & \textbf{64.6}  & \textbf{75.4}  & 81.5 & \textbf{42.9} & 46.8  & 61.0  & 77.9  \\
B-Bd + B-Et &Textual &  42.5 & 47.9  & 60.4  & 78.5 & \textbf{60.0} & \textbf{64.6}  & 73.8  & \textbf{89.2} &37.0 & 41.6  & 53.2  & 71.4  \\  
\hline 

ISNs~\cite{hierarchical}&Visual & 2.5 & 4.4 & 12.0 & 31.0 & \textbf{16.9} & \textbf{26.2} & \textbf{40.0} & \textbf{55.4} & 0.6 & 1.9 & 7.8 & 29.9 \\
CLIP$_i$  &Visual & \textbf{4.6} & \textbf{6.3}  & \textbf{15.4}  & \textbf{41.1} &4.6 & 4.6  & 13.8  & 41.5 &5.2 & \textbf{8.4}  & \textbf{19.5}  & 42.2   \\
Lo + Ob  &Visual & 3.2 & 3.7  & 8.5  & 27.7 &10.8 & 10.8  & 13.8  & 27.7 &3.9 & 4.5  & 9.7  & 27.3    \\
Sc + Ob  &Visual & 1.2 & 1.5  & 6.4  & 20.6 &3.1 & 4.6  & 9.2  & 21.5 &1.9 & 2.6  & 9.1  & 26.0   \\
Lo + Sc  &Visual & 2.4 & 3.0  & 9.0  & 27.1 &6.2 & 6.2  & 12.3  & 33.8 &1.9 & 3.2  & 9.1  & 26.0 \\ 
Lo + Sc + Ob  &Visual  & 2.5 & 3.4  & 9.1  & 31.0 &7.7 & 7.7  & 13.8  & 30.8 &3.2 & 5.2  & 11.7  & 33.8  \\ 
Lo + CLIP$_i$  &Visual &  3.7 & 5.6  & 12.9  & 36.9 &4.6 & 7.7  & 13.8  & 41.5 &5.2 & 7.1  & 16.2  & 37.7  \\
Sc + CLIP$_i$  &Visual & 3.9 & 5.6  & 12.7  & 36.4 &3.1 & 4.6  & 13.8  & 44.6 & \textbf{5.8} & \textbf{8.4}  & 16.9  & 40.3 \\
Lo + Sc + CLIP$_i$ &Visual  & 2.5 & 3.7  & 10.5  & 33.8 &4.6 & 4.6  & 10.8  & 33.8 &3.2 & 5.2  & 11.7  & \textbf{43.5} \\ 
\hline

CLIP$_i$ + B-Bd + B-Et & Textual+Visual & 63.6 & 68.9  & 78.5  & 86.6 & 69.2 & \textbf{76.9}  & \textbf{84.6}  & \textbf{90.8} &61.0 & 64.9  & 74.7  & 81.8 \\
CLIP$_i$ + Lo + B-Bd + B-Et & Textual+Visual & 61.4 & 66.0  & 76.8  & 86.3 &66.2 & 70.8  & 81.5  & \textbf{90.8} &61.0 & 64.9  & 74.0  & 81.8\\
CLIP$_i$ + Sc + B-Bd + B-Et & Textual+Visual & 63.1 & 68.0  & 78.0  & 86.0 &63.1 & 67.7  & 76.9  & 86.2 & \textbf{63.6} & 68.8  & 77.3  & 83.8   \\
Lo + Sc + B-Bd + B-Et & Textual+Visual & 65.1 & 69.5  & 78.7  & 84.8 &70.8 & 75.4  & 81.5  & 86.2 & \textbf{63.6} & 68.2  & 77.9  & 80.5 \\
CLIP$_i$ + Lo + Sc + B-Bd + B-Et & Textual+Visual & \textbf{65.5} & \textbf{70.6}  & \textbf{81.2}  & \textbf{88.7} & \textbf{72.3} & \textbf{76.9}  & 83.1  & \textbf{90.8} & \textbf{63.6} & \textbf{69.5}  & \textbf{81.2}  & \textbf{85.7}  \\
\hline
\end{tabular}
}\end{center}
\end{table}

\textbf{Textual Models}: For smaller GCD thresholds specifically city and region, in \textit{T2}, the combination \textit{B-Et + B-Bd} improves the performance, and in \textit{T1} and \textit{T3} the \textit{B-Et} model provides the best results.
When used separately, \textit{B-Et} has a more substantial impact than \textit{B-Bd}, indicating that named entities and their frequency play a vital role to predict the focus location in the news.
%
While \textit{Mordecai} and \textit{Cliff-clavin} achieve the best results at country and continent level for \textit{T1} and \textit{T3}, respectively, these baselines are either not applicable~(\textit{Mordecai}) or achieve worse results~(\textit{Cliff-clavin}) compared to our models on more fine-grained levels. 

\textbf{Visual Models}: 
The results show that \textit{CLIP$_i$} performs well in all test variants providing the best results on T1 and T3, and that combinations with scene~(\textit{Sc + CLIP$_i$}) and location features~(\textit{Lo + Sc + CLIP$_i$}) can further improve the results. ISNs specifically trained for photo geolocalization achieve superior results on \textit{T2} where images depict the query location and provide enough geographical cues. Unlike \textit{CLIP$_i$}, ISNs do not generalize well on other test variants.

\textbf{Multimodal Models}: 
The combination of \textit{CLIP$_i$} with multimodal information drastically improves the results compared to unimodal models in all test data variants and distance thresholds. 
Even though our visual models do not outperform \textit{ISNs} in \textit{T2}, they considerably improve the results when combined with textual features~(\textit{Lo + Sc + B-Bd + B-Et}). 
These results suggest that a multimodal architecture is beneficial for focus location estimation in news.

\begin{figure}[t]
  \includegraphics[width=\textwidth]{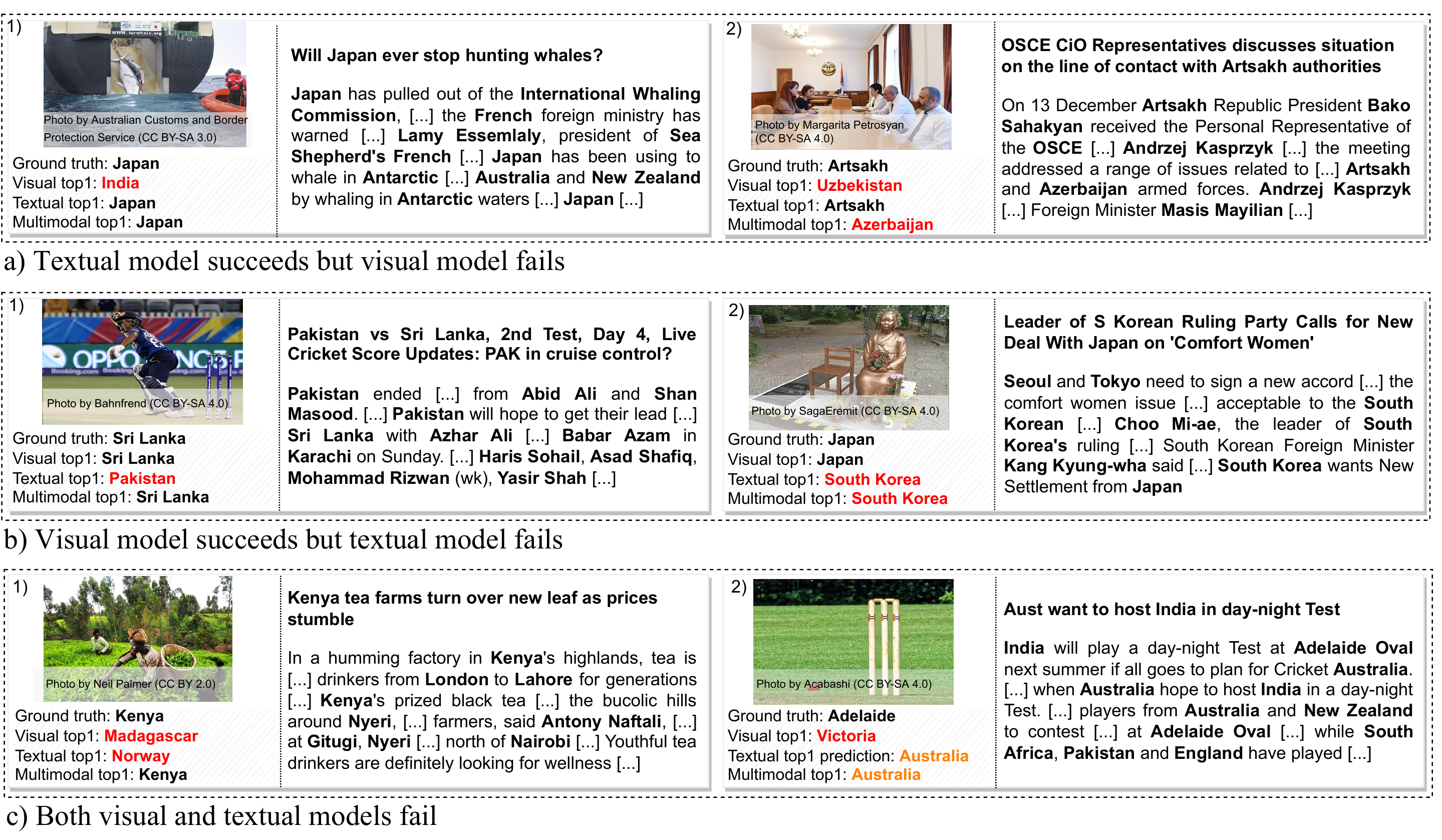}  \caption{Sample outputs of the MM-Locate-News dataset. In the left column the multimodal model predicts the correct location but in the right column it fails. Images are replaced with similar ones due to license restrictions.}
  \label{fig:qualitative_examples}
\end{figure}
\textbf{Qualitative Results}:
Figure~\ref{fig:qualitative_examples} presents the sample outputs for different models. 
As expected, it is easier for the textual model to predict the correct focus location when there are multiple mentions of the ground truth location in the text (Figure~\ref{fig:qualitative_examples}-a.1), or person names related to the focus location such as Figure~\ref{fig:qualitative_examples}-a.2, which mentions \textit{Masis Mayilian} the forign minister of \textit{Artsakh}. 
However, the textual model fails when there are multiple locations (Figure~\ref{fig:qualitative_examples}-b.2) or person names (Figure~\ref{fig:qualitative_examples}-b.1) which are irrelevant to the focus location.
It is very challenging for the visual model to predict the correct focus location due to lack of visual geo-representative features in news images (Figure~\ref{fig:qualitative_examples}-a.1, \ref{fig:qualitative_examples}-c.2) specifically when the image is indoor (Figure~\ref{fig:qualitative_examples}-a.2). 
In contrast, the image sometimes provides rich content as displayed in Figure~\ref{fig:qualitative_examples}-b.1, which shows the sport competition mentioned in the text, or Figure~\ref{fig:qualitative_examples}-b.2 which illustrates the \textit{Comfort Women statue}, referring to \textit{Japan}.
The multimodal model fails when there are multiple mentions of irrelevant entities and the image lacks geo-representative features such as illustrated in  Figure~\ref{fig:qualitative_examples}-a.2.
In Figure~\ref{fig:qualitative_examples}-c.1, both unimodal models fail but the combination of visual and textual information helps the multimodal model succeed since the text is about the event shown in the picture which is related to the focus location \textit{Kenya}.
Another example is Figure~\ref{fig:qualitative_examples}-c.2, where the image contains scene information (Cricket field) but it does not provide visual information about the event (Cricket sport competition). 
Therefore, the multimodal model fails to predict the city \textit{Adelaide}, but it predicts the correct focus country \textit{Australia}. 
Even though the image provides rich visual features when combined with irrelevant locations in the text (e.g., \textit{South Korea}) the model gets confused.
This demonstrates how much entities in text are important for multimodal focus location estimation. 
To sum up, the qualitative examples confirm the need for both image and text to predict the focus location of news.

\subsection{Experimental Results on BreakingNews}
\label{sec:res_breakingnews}
We also evaluate our proposed focus location estimation architectures on Breaking\-News~\cite{breakingnews} for comparison. We divide the dataset into \num{33376}, \num{11210}, and \num{10581} samples for train, validation, and test splits. 
However, please note that the quality of ground truth labels in Breaking\-News is much lower compared to our proposed MM-Locate-News dataset as the test data are not annotated manually for focus locations. 
For a better comparison to the approaches presented by Ramisa et al.~\cite{breakingnews}, we train a regression model by minimizing the GCD between the ground truth (the first geo-coordinate as mentioned in the original paper~\cite{breakingnews}) and the predicted geographical coordinates based on the various types of state-of-the-art visual and textual features presented in Section~\ref{sec:baseline_models}.

\begin{table}[t]
\begin{center}
\caption{Mean and median GCD (km) using the best models per modality on the BreakingNews test set (mean and median values are divided by 1000).} 
\label{tab:eval_breakingnews}

\begin{tabular}{  l |  c c  } \hline 

\textbf{Approach}   & \textbf{Mean} & \textbf{Median}  \\ \hline

CLIP$_i$               &  2.84 & 1.3 \\ 
B-Bd +  B-Et    & 1.93 & 0.88 \\ 

CLIP$_i$  + Lo + Sc + B-Bd + B-Et  & \textbf{1.88} & \textbf{0.83} \\   
\hline
\hline
Places~\cite{breakingnews}    &     3.40 & \textbf{0.68}\\ 
W2V matrix~\cite{breakingnews} &    1.92 & 0.90 \\ 
VGG19 + Places +
W2V~\cite{breakingnews}&\textbf{1.91} & 0.88 \\ \hline
\end{tabular}
\end{center}
\end{table}

Table~\ref{tab:eval_breakingnews} shows the comparison of the regression models provided by Ramisa et al.~\cite{breakingnews} with our approaches using Mean and Median GCD values (divided by 1000). 
From our proposed models, as expected the multimodal model \textit{CLIP$_i$ + Lo + Sc + B-Bd + B-Et} achieves the best performance. In comparison with \textit{VGG19 + Places + W2V matrix} from Ramisa et al.~\cite{breakingnews}, our model achieves better results with a difference of 30\,km and 50\,km in terms of mean and median GCD, respectively. 
Regarding the proposed unimodal models \textit{B-Bd +  B-Et} outperforms \textit{CLIP$_i$} in terms of both mean and median. 
The proposed textual model \textit{B-Bd + B-Et} achieves slightly better results in comparison with \textit{W2V matrix~\cite{breakingnews}} regarding the median value (with difference of 20\,km). 
Regarding the visual models, CLIP$_i$ outperforms Places~\cite{breakingnews} in terms of mean value by 560\,km. 
However, in terms of median value it is worse with a margin of 620\,km. 
This may be explained by more (bad) outlier predictions of the Places model. 
Overall, the results suggest that a multimodal architecture is beneficial for the task of location estimation in news.


\section{Conclusion}
\label{sec:conc}
In this paper, we have introduced a novel dataset called \textit{MM-Locate-News}, which is a multimodal collection of image-text pairs for focus location estimation of news articles. 
A weakly-supervised method has been suggested to collect news with focus locations. 
We manually annotated the test data to acquire three different test data variants.
We have proposed various baselines and multimodal approaches using state-of-the-art transformer-based architectures for the task of focus location estimation.
The experimental results have shown that the combination of textual and visual features for geolocation estimation outperforms the compared approaches that rely on features from a single modality. 

In future work, we will extend the dataset by considering additional languages and locations. We will also investigate multimodal geolocation estimation on different news domains and apply further noise removal to the images.
\\
\\
\textbf{Acknowledgements.} This work was partially funded by the EU Horizon 2020 research and innovation program under the Marie Skłodowska-Curie grant agreement no.~812997 (CLEOPATRA ITN), and by the Ministry of Lower Saxony for Science and Culture (Responsible AI in digital society, project no.~51171145).

%

\bibliographystyle{splncs04}
\bibliography{references}

\begin{thebibliography}{10}
\providecommand{\url}[1]{\texttt{#1}}
\providecommand{\urlprefix}{URL }
\providecommand{\doi}[1]{https://doi.org/#1}

\bibitem{andogah2012every}
Andogah, G., Bouma, G., Nerbonne, J.: Every document has a geographical scope.
  Data \& Knowledge Engineering  \textbf{81},  1--20 (2012)

\bibitem{mlm}
Armitage, J., Kacupaj, E., Tahmasebzadeh, G., Swati, Maleshkova, M., Ewerth,
  R., Lehmann, J.: {MLM:} {A} benchmark dataset for multitask learning with
  multiple languages and modalities. In: International Conference on
  Information and Knowledge Management, {CIKM}. pp. 2967--2974 (2020),
  \url{https://doi.org/10.1145/3340531.3412783}

\bibitem{wikifier}
Brank, J., Leban, G., Grobelnik, M.: Semantic annotation of documents based on
  wikipedia concepts. Informatica (Slovenia)  \textbf{42}(1) (2018),
  \url{http://www.informatica.si/index.php/informatica/article/view/2228}

\bibitem{DBLP:conf/www/CrandallBHK09}
Crandall, D.J., Backstrom, L., Huttenlocher, D.P., Kleinberg, J.M.: Mapping the
  world's photos. In: International Conference on World Wide Web, {WWW}. pp.
  761--770 (2009), \url{https://doi.org/10.1145/1526709.1526812}

\bibitem{deng2009imagenet}
Deng, J., Dong, W., Socher, R., Li, L.J., Li, K., Fei-Fei, L.: Imagenet: A
  large-scale hierarchical image database. In: IEEE Conference on Computer
  Vision and Pattern Recognition. pp. 248--255 (2009)

\bibitem{bert}
Devlin, J., Chang, M., Lee, K., Toutanova, K.: {BERT:} pre-training of deep
  bidirectional transformers for language understanding. In: Conference of the
  North American Chapter of the Association for Computational Linguistics:
  Human Language Technologies, {NAACL-HLT}. pp. 4171--4186 (2019),
  \url{https://doi.org/10.18653/v1/n19-1423}

\bibitem{clavin}
D'Ignazio, C., Bhargava, R., Zuckerman, E., Beck, L.: Cliff-clavin: Determining
  geographic focus for news articles. In: NewsKDD:Data Science for News
  Publishing Workshop co-located with {ACM} SIGKDD Conference on Knowledge
  Discovery and Data Mining (2014)

\bibitem{DBLP:conf/eccv/GordoARL16}
Gordo, A., Almaz{\'{a}}n, J., Revaud, J., Larlus, D.: Deep image retrieval:
  Learning global representations for image search. In: European Conference on
  Computer Vision, {ECCV}. pp. 241--257 (2016),
  \url{https://doi.org/10.1007/978-3-319-46466-4\_15}

\bibitem{geovirus}
Gritta, M., Pilehvar, M.T., Collier, N.: Which melbourne? augmenting geocoding
  with maps. In: 56th Annual Meeting of the Association for Computational
  Linguistics, {ACL}. pp. 1285--1296 (2018),
  \url{https://aclanthology.org/P18-1119/}

\bibitem{mordecai}
Halterman, A.: Mordecai: Full text geoparsing and event geocoding. The Journal
  of Open Source Software  \textbf{2}(9) (2017). \doi{10.21105/joss.00091}

\bibitem{Hays2008}
Hays, J., Efros, A.A.: {IM2GPS:} estimating geographic information from a
  single image. In: Conference on Computer Vision and Pattern Recognition,
  {CVPR} (2008)

\bibitem{resnet}
He, K., Zhang, X., Ren, S., Sun, J.: Identity mappings in deep residual
  networks. In: European Conference on Computer Vision, {ECCV}. pp. 630--645
  (2016), \url{https://doi.org/10.1007/978-3-319-46493-0\_38}

\bibitem{spacy}
Honnibal, M., Montani, I.: {spaCy 2: Natural language understanding with bloom
  embeddings, convolutional neural networks and incremental parsing} (2017),
  \url{https://spacy.io}

\bibitem{profile}
Imani, M.B., Chandra, S., Ma, S., Khan, L., Thuraisingham, B.M.: Focus location
  extraction from political news reports with bias correction. In:
  International Conference on Big Data, {BigData}. pp. 1956--1964 (2017),
  \url{https://doi.org/10.1109/BigData.2017.8258141}

\bibitem{mvmf}
Izbicki, M., Papalexakis, E.E., Tsotras, V.J.: Exploiting the earth's spherical
  geometry to geolocate images. In: European Conference on Machine Learning and
  Knowledge Discovery in Databases, {ECML} {PKDD}. pp. 3--19 (2019),
  \url{https://doi.org/10.1007/978-3-030-46147-8\_1}

\bibitem{kim2017learned}
Kim, H.J., Dunn, E., Frahm, J.: Learned contextual feature reweighting for
  image geo-localization. In: Conference on Computer Vision and Pattern
  Recognition, {CVPR}. pp. 3251--3260 (2017),
  \url{https://doi.org/10.1109/CVPR.2017.346}

\bibitem{kordopatis2017geotagging}
Kordopatis{-}Zilos, G., Papadopoulos, S., Kompatsiaris, I.: Geotagging text
  content with language models and feature mining. Proceedings of the IEEE
  \textbf{105}(10),  1971--1986 (2017),
  \url{https://doi.org/10.1109/JPROC.2017.2688799}

\bibitem{placing_language_model}
Kordopatis{-}Zilos, G., Popescu, A., Papadopoulos, S., Kompatsiaris, Y.:
  Placing images with refined language models and similarity search with
  pca-reduced {VGG} features. In: MediaEval 2016 Workshop. vol.~1739 (2016),
  \url{http://ceur-ws.org/Vol-1739/MediaEval\_2016\_paper\_13.pdf}

\bibitem{krippendorff2011computing}
Krippendorff, K.: Computing krippendorff's alpha-reliability  (2011),
  \url{https://repository.upenn.edu/asc_papers/43}

\bibitem{kulkarni2021multi}
Kulkarni, S., Jain, S., Hosseini, M.J., Baldridge, J., Ie, E., Zhang, L.:
  Multi-level gazetteer-free geocoding. In: International Workshop on Spatial
  Language Understanding and Grounded Communication for Robotics. pp. 79--88
  (2021)

\bibitem{mediaeval2}
Larson, M.A., Soleymani, M., Gravier, G., Ionescu, B., Jones, G.J.F.: The
  benchmarking initiative for multimedia evaluation: Mediaeval 2016. IEEE
  MultiMedia  \textbf{24}(1),  93--96 (2017),
  \url{https://doi.org/10.1109/MMUL.2017.9}

\bibitem{lgl}
Lieberman, M.D., Samet, H., Sankaranarayanan, J.: Geotagging with local
  lexicons to build indexes for textually-specified spatial data. In:
  International Conference on Data Engineering, {ICDE}. pp. 201--212 (2010),
  \url{https://doi.org/10.1109/ICDE.2010.5447903}

\bibitem{DBLP:conf/cvpr/LinBH13}
Lin, T., Belongie, S.J., Hays, J.: Cross-view image geolocalization. In:
  Conference on Computer Vision and Pattern Recognition {CVPR}. pp. 891--898
  (2013), \url{https://doi.org/10.1109/CVPR.2013.120}

\bibitem{hierarchical}
M{\"{u}}ller{-}Budack, E., Pustu{-}Iren, K., Ewerth, R.: Geolocation estimation
  of photos using a hierarchical model and scene classification. In: European
  Conference on Computer Vision, {ECCV}. pp. 575--592 (2018),
  \url{https://doi.org/10.1007/978-3-030-01258-8\_35}

\bibitem{DBLP:journals/ijmir/Muller-BudackTD21}
M{\"{u}}ller{-}Budack, E., Theiner, J., Diering, S., Idahl, M., Hakimov, S.,
  Ewerth, R.: Multimodal news analytics using measures of cross-modal entity
  and context consistency. International Journal of Multimedia Information
  Retrieval  \textbf{10}(2),  111--125 (2021),
  \url{https://doi.org/10.1007/s13735-021-00207-4}

\bibitem{clip}
Radford, A., Kim, J.W., Hallacy, C., Ramesh, A., Goh, G., Agarwal, S., Sastry,
  G., Askell, A., Mishkin, P., Clark, J., Krueger, G., Sutskever, I.: Learning
  transferable visual models from natural language supervision. In:
  International Conference on Machine Learning, {ICML}. pp. 8748--8763 (2021),
  \url{http://proceedings.mlr.press/v139/radford21a.html}

\bibitem{breakingnews}
Ramisa, A., Yan, F., Moreno{-}Noguer, F., Mikolajczyk, K.: Breakingnews:
  Article annotation by image and text processing. IEEE Transactions in Pattern
  Analysis and Machine Intelligence  \textbf{40}(5),  1072--1085 (2018),
  \url{https://doi.org/10.1109/TPAMI.2017.2721945}

\bibitem{cplanet}
Seo, P.H., Weyand, T., Sim, J., Han, B.: Cplanet: Enhancing image
  geolocalization by combinatorial partitioning of maps. In: European
  Conference on Computer Vision, {ECCV}. pp. 544--560 (2018),
  \url{https://doi.org/10.1007/978-3-030-01249-6\_33}

\bibitem{DBLP:conf/sigir/SerdyukovMZ09}
Serdyukov, P., Murdock, V., van Zwol, R.: Placing flickr photos on a map. In:
  SIGIR Conference on Research and Development in Information Retrieval,
  {SIGIR}. pp. 484--491 (2009), \url{https://doi.org/10.1145/1571941.1572025}

\bibitem{event_data}
Ulfelder, J., Schrodt, P.: Political instability task force worldwide
  atrocities event data collection codebook. version 1.0 b2 (2009)

\bibitem{BurakEvan}
Uzkent, B., Sheehan, E., Meng, C., Tang, Z., Burke, M., Lobell, D.B., Ermon,
  S.: Learning to interpret satellite images using wikipedia. In: International
  Joint Conference on Artificial Intelligence, {IJCAI}. pp. 3620--3626 (2019),
  \url{https://doi.org/10.24963/ijcai.2019/502}

\bibitem{DBLP:journals/cacm/VrandecicK14}
Vrandecic, D., Kr{\"{o}}tzsch, M.: Wikidata: a free collaborative
  knowledgebase. Communications of the ACM  \textbf{57}(10),  78--85 (2014),
  \url{https://doi.org/10.1145/2629489}

\bibitem{googlelandmarkdataset}
Weyand, T., Araujo, A., Cao, B., Sim, J.: Google landmarks dataset v2 - {A}
  large-scale benchmark for instance-level recognition and retrieval. In:
  Conference on Computer Vision and Pattern Recognition, {CVPR}. pp. 2572--2581
  (2020)

\bibitem{zhouplaces}
Zhou, B., Lapedriza, {\`{A}}., Khosla, A., Oliva, A., Torralba, A.: Places: {A}
  10 million image database for scene recognition. IEEE Transactions on Pattern
  Analysis and Machine Intelligence  \textbf{40}(6),  1452--1464 (2018),
  \url{https://doi.org/10.1109/TPAMI.2017.2723009}

\end{thebibliography}

\end{document}